\documentclass[conference]{IEEEtran}

\IEEEoverridecommandlockouts
\usepackage{cite}
\usepackage{amsmath,amssymb,amsfonts}
\usepackage{algorithmic}
\usepackage{graphicx}
\usepackage{textcomp}
\usepackage{xcolor}
\usepackage{amsmath}
\usepackage{algorithm,algorithmic}
\usepackage{cellspace}    
\usepackage{makecell, multirow, tabularx}
\usepackage{subcaption,graphicx}
\usepackage{pbox}
\usepackage[british]{babel}
\usepackage{array} 
\usepackage{float}

 \usepackage{balance}
\balance

\usepackage[skip=1pt]{caption}
\usepackage{enumitem}

\def\BibTeX{{\rm B\kern-.05em{\sc i\kern-.025em b}\kern-.08em
    T\kern-.1667em\lower.7ex\hbox{E}\kern-.125emX}}

    \usepackage{mathtools}

\usepackage[top=0.77in,bottom=1.53in,left=0.65in,right=0.65in]{geometry}

\begin{document}

\title{Balancing Subjectivity and Objectivity in Network Selection: A Decision-Making Framework Towards Digital Twins}

\DeclareRobustCommand{\IEEEauthorrefmark}[1]{\smash{\textsuperscript{\footnotesize #1}}}

\author{
    \IEEEauthorblockN{Brahim Mefgouda$^{1,2}$, Hanen Idoudi$^{3}$,   Mohammad Al-Quraan$^{4}$, Ismail Lotfi$^{1,5}$, Omar Alhussein$^{1,5}$, \\ Lina Mohjazi$^{4}$, and Sami Muhaidat$^{1,2,6}$}
    \IEEEauthorblockA{$^{1}$KU 6G Research Center, Khalifa University, Abu Dhabi, UAE\\
                      $^{2}$Department of Communication and Information Systems, Khalifa University, Abu Dhabi, UAE\\
                      $^{3}$King Faisal University (KFU), Al Ahsa, Saudi Arabia\\
                      $^{4}$ James Watt School of Engineering, University of Glasgow, 
G12 8QQ Glasgow, U.K\\
                      $^{5}$Department of Computer Science, Khalifa University, Abu Dhabi, UAE\\
                      $^{6}$Department of Systems and Computer Engineering, Carleton University, Ottawa, Canada\\}

                      Emails: \{brahim.mefgouda, ismail.lotfi, omar.alhussein, sami.muhaidat\}@ku.ac.ae, hidoudi@kfu.edu.sa,\\ \{mohammad.alquraan, lina.mohjazi\}@glasgow.ac.uk
                      
                                            }


\vspace{-2mm}

\maketitle
\begin{abstract}

Selecting the optimal radio access technology (RAT) during vertical handovers (VHO) in heterogeneous wireless networks (HWNs) is critical. Multi-attribute decision-making (MADM) is the most common approach used for network selection (NS) in HWNs. However, existing MADM-NS methods face two major challenges: the rank reversal problem (RRP), where the relative ranking of alternatives changes unexpectedly, and inefficient handling of user and/or service requirements. These limitations result in suboptimal RAT selection and diminished quality of service, which becomes particularly critical for time-sensitive applications. To address these issues, we introduce in this work a novel weighting assignment technique called BWM-GWO, which integrates the Best-Worst Method (BWM) with the Grey Wolf Optimization (GWO) algorithm through a convex linear combination.  The proposed framework achieves a balanced decision-making process by using BWM to compute subjective weights that capture user/service preferences, while employing GWO to derive objective weights aimed at minimizing RRP.  The development and validation of this framework establish a digital model for NS in HWNs, marking the initial step toward realizing a digital twin (DT). Experimental results show that integrating the proposed BWM-GWO technique with MADM-NS reduces RRP occurrence by up to 71.3\% while significantly improving user and service satisfaction compared to benchmark approaches.





\end{abstract}

\begin{IEEEkeywords}
Network selection, heterogeneous wireless networks,  vertical handover, grey wolf optimization, decision making, digital twins.
\end{IEEEkeywords}
\thispagestyle{plain}
\pagestyle{plain}

\section{Introduction}
\subsection{Background}
Over the last decade, the evolution of various radio access technologies (RATs), including IEEE standards (e.g., WiMAX, Wi-Fi) and cellular networks (e.g., 4G, 5G), alongside the equipping of new mobile devices with multiple access network interfaces, have led to the emergence of heterogeneous wireless networks (HWNs)~\cite{xu2021survey}.
While HWNs provide numerous benefits to user equipment (UE), including seamless access to RATs anywhere and anytime, the network selection (NS) problem remains a critical challenge. This problem requires the UE to consistently connect to the most suitable RAT that aligns with user and/or service requirements~\cite{wang2012mathematical}.

Numerous mathematical approaches have been proposed in the literature to tackle the NS problem in HWNs, including multi-attribute decision-making (MADM), neural networks, and fuzzy logic (FL). Among these, MADM approaches have gained prominence as the preferred solution due to their distinct advantages, such as real-time decision-making capabilities, ease of implementation, computational efficiency, and superior scalability compared to other methods~\cite{abdullah2024heterogeneous,wang2012mathematical}. However, MADM-based NS (MADM-NS) methods face two significant limitations: \textit{(i)} their tendency to select networks based solely on high aggregate scores while disregarding specific user and service requirements;  \textit{(ii)} their susceptibility to the rank reversal problem (RRP), where network rankings change unexpectedly upon addition or removal of alternatives. These limitations result in suboptimal NS decisions, triggering unnecessary handovers that generate substantial signalling overhead and temporary service interruptions. {\color{black}   In 6G networks, these limitations become more critical due to increased heterogeneity and ultra-dense deployments. The inefficiencies of MADM-NS can compromise key 6G performance metrics, leading to frequent handovers, increased signaling overhead, and degraded quality of service (QoS),  impacting delay-sensitive applications such as immersive video streaming and mission-critical communications~\cite{mefgouda2024qos,ma2021intelligent,Quraan}.}

A promising approach to overcome the NS challenge involves leveraging the capabilities of digital twin (DT) technology.  DT is a dynamic, real-time virtual representation of physical systems, which evolves from a static digital model to a digital shadow and ultimately into a fully interactive digital model~\cite{Ismail_2023_JSAC}. To date, limited studies have explored the use of DTs for solving the NS problem, such as~\cite{zheng2022digital}, which explores using DTs to optimize HWNs selection through a Markov Decision Process-based approach. {\color{black}In contrast, our work addresses the NS problem by developing a DT-driven framework that optimizes MADM-NS. We aim to develop a DT framework to optimize the MADM-NS problem, starting with a foundational digital model and progressively evolving it into a fully functional DT.}




{\color{black}

}

\subsection{Related Work}

{\color{black}
Various solutions have been proposed in the literature to solve MADM-NS drawbacks, which can be classified into two main categories: normalization-based and weighting-based. The former includes a focus on altering the original normalization technique and choosing a more appropriate substitute. However, The latter optimizes the attribute weight assignments for RATs. Specifically, instead of relying on static or predefined weights, these solutions leverage advanced optimization techniques to dynamically determine the most appropriate attribute weights.

Normalization-based solutions are based on the theory that appropriate normalization methods can address MADM-NS limitations~\cite{jahan2015state}.  MADM-NS approaches use normalization to eliminate unit differences (e.g., money or time) among criteria. However, inconsistencies in normalization techniques can alter normalized values despite unchanged original values, impacting scores and final rankings. To overcome this issue, various approaches, such as that of M. A. Senouci \emph{et al.}~\cite{senouci2016topsis} decreased the RRP by replacing the original normalization method of Technique for Order of Preference by Similarity to Ideal Solution (TOPSIS) with max-min normalization. The work in~\cite{senouci2016utility} successfully addressed the RRP in TOPSIS by employing utility functions (UFs).  In~\cite{mansouri2019new}, the RRP was eliminated and the number of VHOs was reduced using FL to standardize the decision matrix using the manhattan distance method. In our previous work~\cite{mefgouda2024qos}, we avoided RRP and achieved the required QoS for streaming applications in the measurement of alternatives and ranking according to the compromise solution (MARCOS) by introducing a novel sigmoid UFs. However, MADM-NS benchmarks, including TOPSIS and simple additive weighting (SAW), outperformed our approach in meeting QoS requirements due to their consideration of subjectivity. This limitation arises because our method focuses solely on normalization to mitigate RRP (objectivity), without accounting for subjectivity in NS.

}

In the second class, several approaches have been proposed, including those in~\cite{almutairi2018genetic, almutairi2021particle,mefgouda2021multi, mefgouda2023new}, which leverage metaheuristic algorithms to address the constraints of MADM-NS by optimizing the summation of absolute differences in ranking values among candidate RATs. While these solutions effectively mitigate (and in many cases eliminate) the RRP, they often fail to efficiently accommodate user and service preferences (subjectivity). Similarly, the work in~\cite{radouche2021new} reduced the RRP using the particle swarm optimization algorithm. However, the challenge of adequately addressing user and service requirements persists. Other solutions focused on combining objectivity with subjectivity to reduce the RRP while simultaneously achieving the user/service requirements, respectively. For instance,~\cite{priya20205gaunets} combined the fuzzy analytic hierarchy process (AHP) with TOPSIS to optimize decision weights, while~\cite{mansouri2020use} incorporated FL with TOPSIS to better address user preferences. Similarly,~\cite{yu2020novel} combined coefficient of variation for objective weights and intuitionistic normal fuzzy AHP for subjective weights. 
\vspace{-0.4em}
\subsection{Contribution}
Choosing an appropriate MADM normalization technique can help mitigate the RRP. However, these solutions often fall short of aligning with user and/or service preferences. Enhancing the MADM-NS through decision criteria weighting can address some of its shortcomings. While metaheuristics algorithms can reduce the RRP, they may still struggle to meet user/service preferences effectively. Conversely, approaches using different weighting techniques, such as AHP, align well with user/service preferences but continue to face challenges with the RRP.

In this work, we propose a novel weighting assessment technique called (BWM-GWO), which integrates the best-worst method (BWM) with the grey wolf optimization (GWO) algorithm through a convex linear equation. This hybrid approach is designed to seamlessly enhance any MADM-NS framework by addressing its inherent limitations. The BWM component introduces subjectivity by emphasizing user and service evaluations, while the GWO component ensures objectivity by mitigating the RRP. The convex linear combination serves to balance these dual objectives, enabling an effective trade-off between user/service preferences and methodological rigor.  {\color{black}To validate the proposed framework, we evaluate its performance within a digital model of HWNs, paving the way for the development of DTs. }


The GWO is a metaheuristic algorithm inspired by the hunting behaviour and social hierarchy of grey wolves~\cite{faris2018grey}. We apply the GWO algorithm for its demonstrated superiority over other metaheuristics, attributed to its ability to maintain an optimal balance between exploration and exploitation~\cite{mirjalili2014grey}. This balance enables GWO to achieve faster convergence and deliver higher-quality solutions across various complex optimization problems.  In parallel, the BWM is a MADM technique that calculates the criteria weights by comparing the best and worst criteria relative to all others~\cite{rezaei2015best}. The use of BWM for subjective weight calculation represents a novel application in the network selection domain. To the best of our knowledge, the adoption of BWM for subjective weight calculation is particularly novel, as it has not been previously applied in the NS field. In addition, BWM demonstrates superiority over traditional MADM weighting assessment techniques, further underscoring its potential to provide robust and effective solutions within the MADM-NS framework.




\begin{figure*}[h]
    \centering
    \begin{minipage}{0.25\textwidth}
        \centering
        \includegraphics[width=\linewidth]{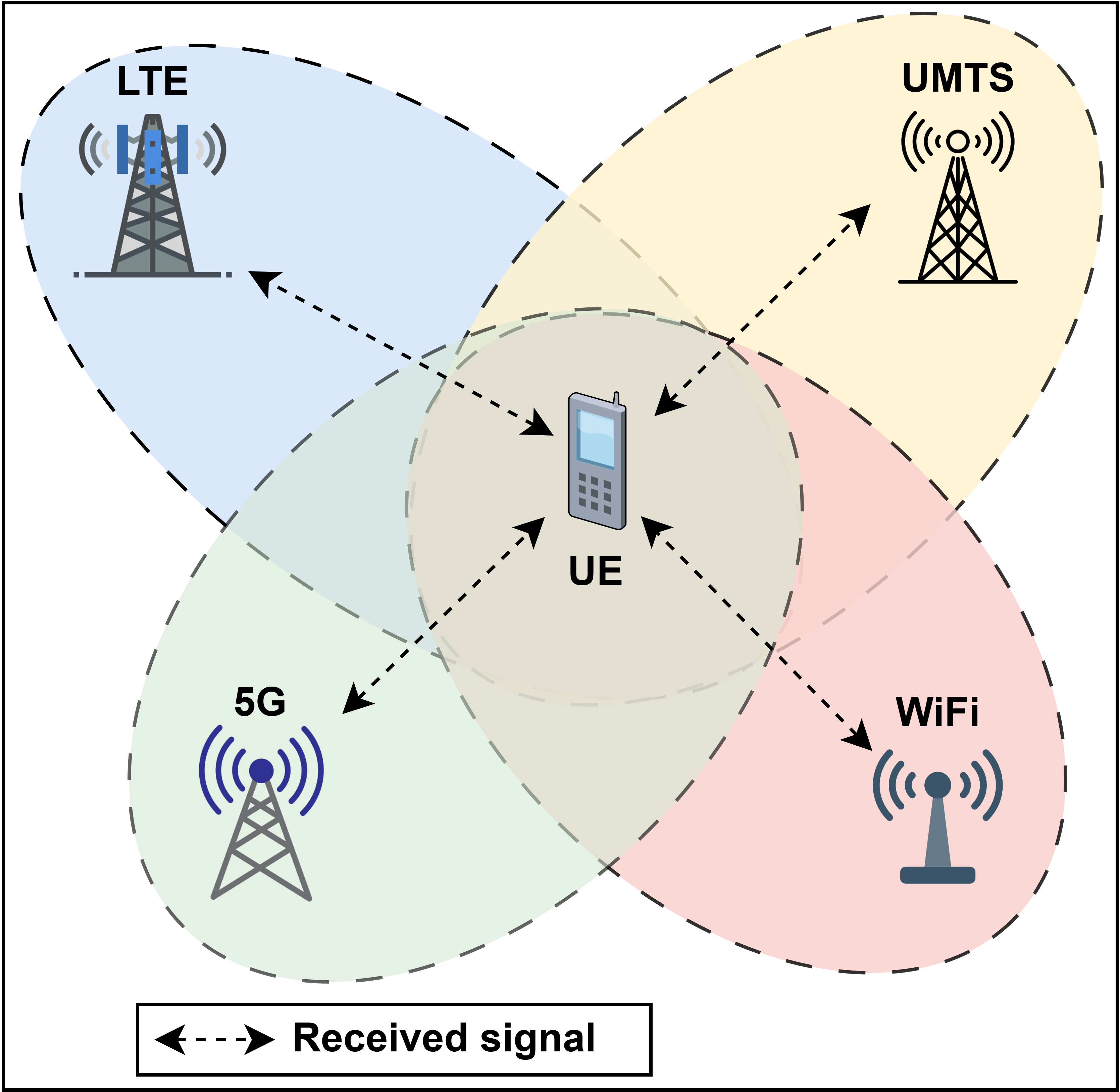} 
        \caption{HWNs environment.}
        \label{fig:System}
    \end{minipage}
    \hfill
    \begin{minipage}{0.72\textwidth}
        \centering
        \includegraphics[width=\linewidth]{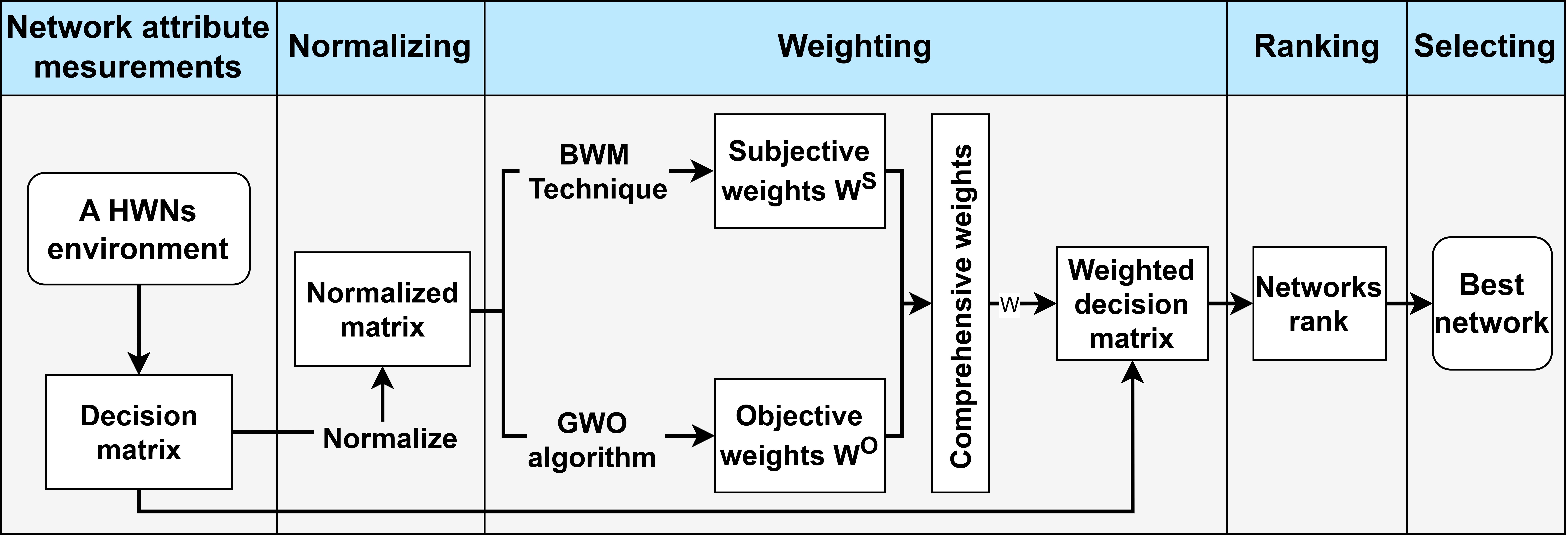} 
        \caption{General architecture of using BWM-GWO in MADM-NS.}
        \label{fig:General Architecture}
    \end{minipage}
    \vspace{-3mm}
\end{figure*}


\section{System Model and Problem formulation}
\subsection{System Overview}

We considered an outdoor HWNs environment composed of multiple overlapping RATs, including WiFi, WiMAX, LTE, and 5G, as illustrated in Fig.~\ref{fig:System}.  WiFi leverages unlicensed 2.4GHz and 5GHz bands with CSMA/CA to offer high data rate (DR). WiMAX, operating below 11GHz, uses OFDMA for wireless metropolitan coverage and mobility support. LTE, in licensed bands between 700MHz and 2600MHz, employs OFDMA and SC-FDMA to provide wide-area connectivity. Finally, 5G extends these capabilities via sub-6GHz and mmWave bands, massive MIMO, and advanced beamforming, enabling ultra-high DR, low latency, and enhanced reliability. We also consider multi-homed UE devices, each capable of connecting to only one wireless network for any given session at a time. As the UE roams in the HWNs, it consistently remains under the coverage of all four types of RATs. Since the primary focus is on selecting the optimal RAT, we omitt explicit considerations of UE speed and the received signal strength.

The VHO process consists of three phases: (i) Collecting network information used to select the best RAT;  (ii) the NS phase; (iii) VHO execution to build the communication link between the chosen RAT and the UE. In our system, each UE is supposed to use the IEEE 802.21 standard~\cite{taniuchi2009ieee} to maintain seamless connectivity in HWNs. This standard identifies available RATs and gathers details through the media independent information service (MIIS). After choosing the optimal RAT, it leverages the Media Independent Command Service (MICS) to handle connections and handovers. The UEs can detect the RATs available within HWNs and retrieve their associated attributes through the MIH standard. Each RAT is defined by six attributes: cost per byte (CB), security (S), DR, packet delay (D), packet jitter (J), and packet loss rate (PLR). 
\subsection{MADM for NS Problem}
We consider an HWNs consisting of $N$ RATs, each defined by $M$ attributes. The MADM approaches are modelled to solve the NS problem as follows:
\begin{enumerate}
\item \textbf{Creating the initial Decision Matrix (DM):} 
This matrix represents the HWNs environment, with rows for available RATs and columns for their attributes.
    \item \textbf{Normalizing the DM:} The DM is normalized to eliminate unit and scale differences among decision criteria.
    \item  \textbf{Calculating the RAT attribute weights vector:} This vector represents the importance of each criterion, with higher weights indicating greater significance. It is expressed as:
    \begin{align}
        W = \{w_i \mid i = 1, 2, \ldots, M\},
        \quad \text{with} \quad \sum_{i=1}^{M} w_i = 1,
    \end{align}
    \noindent where $w_i$ is the weight assigned to the $i^\text{th}$ criterion.

    
    \item \textbf{Ranking the available RATs:} The MADM methods employ different approaches to rank the available RATs based on their performance across the weighted criteria.   

\end{enumerate}

\subsection{Subjective-Objective Weighting Assessment Technique}



In MADM-NS, AHP remains the most widely adopted method \cite{abdullah2024heterogeneous}. In our work, we introduce a novel approach that reformulates the determination of RAT attribute weights as a constrained optimization problem. We redefine the determination of RAT attribute weights as a constrained optimization problem. Our methodology seamlessly integrates both subjective and objective factors. The proposed formulation consists of two key components, as outlined below:
\vspace{1mm}
        \begin{align}\footnotesize	
    W_j = \alpha . W^S_j + \beta . W^O_j,
        \end{align}

with: 
                        \begin{align}\footnotesize	
    \sum\nolimits_{i=i}^{M} W_{i} = 1 \text{ and } \sum\nolimits_{i=i}^{M} 0 \leq W_{i} \leq 1,
                        \label{constr5}
                        \end{align}

where $W$ is the weighting vector of the NS problem derived by integrating the subjective weights ($W^S_j$) obtained through the BWM-NIS with the objective weights ($W^O_j$) obtained through the GWO-NIS. Meanwhile,  $\alpha$ and $\beta$ represent the proportions of subjective and objective weights, respectively, within the composite weighting framework.

\subsubsection{\textbf{Subjective Weights}}
\label{impo}




Determining the best and worst decision criteria for the NS problem requires considering the type of service. In this context, we consider the four QoS service classes defined by the $3^{rd}$ generation partnership project (3GPP)~\cite{mefgouda2023new}: conversational, background, interactive, and streaming. These classes can provide valuable guidance in choosing the best/worst decision criteria for the NS problem. 

The UE defines a subjective importance list, denoted as \( {Limpo} \), to prioritize decision criteria for a chosen traffic class. This list reflects the subjective weighting of attributes based on the specific requirements of the traffic class and incorporates the UE’s experiential insights. Structured hierarchically with length \( M \) (\( |{Limpo}| = M \)), the list ranks criteria by importance, where \( {Limpo}_i \) is more critical than \( {Limpo}_{i+x} \) for \( 1 \leq x \leq M-i \). The BWM technique utilizes this list to compute \( W^S_j \).

\subsubsection{\textbf{Objective Weights}}
We model the objective weights ($W^o_j$) of the NS field as a constrained optimization problem as follows:

\begin{itemize}
    \item[--] An agent is an array of weights $W_{[M]}$ where  $M$ is the maximum number of the network attributes, and $W_{i},$ denotes the weight of the $ i^{th}$ decision criteria. 
    \item[--] $W_{i, L_{Important}} > W_{i, L_{No-Important}}$, where $L_{Important}$ and $L_{No-Important}$ refer to lists of important and non-important decision criteria for a traffic class that runs on the user equipment, respectively, as outlined in our paper~\cite{mefgouda2023new}. 
\end{itemize}

The metahersitc algorithm is applied to reduce the RRP by emphasizing the summation of the absolute value (SV) of the ranking values differences of the candidate networks. Thus, the objective function used to optimize the objective decision criteria is defined as follows: 
     \begin{center}
     \vspace{-3mm}\footnotesize	
        \begin{align}
                 SV = \sum\nolimits_{i=1}^{N} {     \sum\nolimits_{j=i+1}^{N} {    \lvert  {N_i-N_j} \rvert  }    },
                \label{C1}
        \end{align}
    \end{center} 
\noindent where $N_i$ represents the score of the $i^{\text{th}}$ network. The equation for $N_i$ depends on the MADM-NS method used to rank the networks.
$N_i$ denotes the score for the $i^{\text{th}}$ network. The calculation of $N_i$ is contingent upon the MADM-NS approach employed to rank the networks.
\section{Proposed BWM-GWO Technique}
{\color{black}This section describes the implementation of BWM and GWO to optimize the RATs attribute weights and its integration with the MADM-NS approaches.}
\label{Weighting Methods}
\subsection{BWM Technique}
\label{Best-Worst method}
We apply the BWM technique to find the subjective weights for a  given HWNs environment as fellows: 

\begin{itemize}
    
\item Step 1: Identify the set of decision-making criteria.   
\item Step 2: Defining the $Limpo$ list. 

\item Step 3: Determining the best criterion $B$ and the worst criterion $W$, where ($B = Limpo_1$) and ($W = Limpo_M$).

\item Step 4: Assigning the numerical value to indicate the level of preference for the best criterion over all the other criteria using a scale of 1 to 9. The outcome is a vector:
$A_B = (a_{B1},a_{B2},...,a_{Bn})$, where $a_{Bj}$  indicates the preference of the best criterion $B$ over criterion $j$, where   $a_{BB} =1$.

\item Step 5: Assigning the numerical value to indicate the level of preference for the worst criterion over all the other criteria using a scale of 1 to 9. The outcome is a vector:
$A_W = {(a_{1W},a_{2W},...,a_{nW})}^T$, where $a_{Wj}$  indicates the preference of the worst criterion $W$ over criterion $j$.
\item Step 6: Deriving the ideal weights $(w_1^*, w_2^*, \dots, w_n^*)$, ensuring they satisfy the conditions: for each pair $W_B/W_j$ and $W_j/W_W$, the ideal scenario occurs when $W_B/W_j = a_{Bj}$ and $W_j/W_W = a_{Bj}$. This leads to the following problem formulation~\cite{rezaei2015best}:
\vspace{0.5mm}
\begin{subequations}
\label{eq:optz_1}
\begin{align}
\begin{split}
\min \max_j \{ |w_B-a_{BJ}w_j|, |w_j-a_{jW}w_{W}|\}, \label{eq:MaxA} 
\end{split}\\ 
\begin{split}
\hspace{1cm} s.t. \sum_{j=1}^{N} w_j=1\label{eq:MaxB}, \forall 
  j\in \mathcal{N}, w_j \leq 0.
\end{split}
\end{align}
\end{subequations}
\end{itemize}
\subsection{GWO Algorithm}
In this algorithm, the alpha wolf represents the best solution found so far~\cite{mirjalili2014grey}.  Assuming  $t$ is the iteration number and $X_p$ is a vector refers to the current position of the prey.  The grey wolves’ position is specified by vectors as fellows: 
\vspace{0.5mm}
        \begin{align}\footnotesize	
        \overrightarrow{X}(t+1)= \overrightarrow{X_p}(t) - \overrightarrow{A}.\overrightarrow{D},
        \end{align}
        \begin{align}\footnotesize	
        \overrightarrow{D}= |\overrightarrow{C}.\overrightarrow{C_P}(t) - \overrightarrow{X}(t)|,
        \end{align}
        \vspace{0.5mm}
where $\overrightarrow{A}$ and $\overrightarrow{C}$ are calculated as follows: 
        \begin{align}\footnotesize	
        \overrightarrow{A}= 2 \overrightarrow{a}. \overrightarrow{r_1} - \overrightarrow{a}, and         \overrightarrow{C}= 2. \overrightarrow{r_2}.
        \end{align}
        \begin{align}\footnotesize	
        \overrightarrow{C}= 2. \overrightarrow{r_2}.
        \end{align}
In the above equations $\overrightarrow{r_1}$ and $ \overrightarrow{r_2}$  are random vectors within $[0,1]$, while the $\overrightarrow{a}$ are linearly diminished from 2 to 0. 
As described in~\cite{mirjalili2014grey}, the top three positions in the pack are held by wolves designated as "alpha," "beta," and "gamma." Additionally, it was observed that other members of the pack, such as "omega" wolves, would adjust their own positions to align with those deemed the most effective. 

\subsection{BWM-GWO Technique for TOPSIS-NS and SAW-NS}
\label{contribution}

In this work, we investigate the use of MADM techniques in HWNs for VHO management. Specifically, we focus on studying two MADM methods: TOPSIS and SAW.  Fig. \ref{fig:General Architecture} illustrates the general architecture of the approach introduced in this study.  In the what follow, we describe the main changes we propose to the current MADM architeture to enhance its performacne, namely MADM-BWM-GWO.


\subsubsection{Network Attribute Measurements} 
We construct the DM from the collected data. The DM is a matrix of $N$ rows and $M$ columns representing the number of  RATs and the number of evaluation criteria, where the element $DM_{i j}$ denotes the score of the $i^\textsuperscript{th}$ candidate for the $j^\textsuperscript{th}$ attribute.
\subsubsection{Normalizing} 
The type of normalization technique used is dependent on the MADM used to rank the RATs. The normalization method of TOPSIS is defined as follows: 
\vspace{2mm}
\begin{align}\footnotesize	
NM_{ij}= \frac{ DM_{i j}} {\sqrt{  \sum\nolimits_{i=1}^{n}  { DM_{ij}^{2} }     }},
\end{align}
however, the normalization method of SAW is given as follows: 
    \begin{itemize}
        \item[] \vspace{-1mm}\begin{center}
        \begin{equation}	
              \text{For benefit criteria: }   NM_{ij}= \frac{ DM_{ij}}{ \sum\nolimits_{i=1}^{n}{ DM_{ij}}},
                \label{eq20}
        \end{equation}
    \end{center} 
        \item []
        \begin{center}
        \begin{equation}	
             \text{For cost criteria:   }      NM_{ij}= \frac{ \sum\nolimits_{i=1}^{n}{ DM_{ij}} }{ DM_{ij}},
                \label{eq21}
        \end{equation}
    \end{center} 
\end{itemize}
where $NM_{ij}$ represents the normalized value of  the $i^\textsuperscript{th}$ wireless network for the $j^\textsuperscript{th}$ attribute. 

\subsubsection{Weighting} 

The BWM-GWO is applied to compute the suitable decision criteria RATs. Firstly, the BWM is applied to calculate the subjective weights of the service/user preferences. Secondly, the GWO algorithm is used to obtain the objective weights. Finally, the comprehensive method is implimented to combine the subjective and objective weights.

\paragraph{Subjective Weights} The BWM technique defined in section \ref{Best-Worst method} is implemented. The output of the BWM-NS technique is a vector $W^S_j=[W^S_1,W^S_2,...,W^S_M]$, where $W^S_j$ represents the subjective weight of the $j^th$ decision criteria.

\paragraph{Objective Weights} The objective function equation for TOPSIS is defined in Equation (\ref{C2}) by replacing $N_i$ with the TOPSIS coefficient, while the equation for SAW is defined in Equation (\ref{C3}) by substituting $N_i$ with the SAW scoring equation. The objective function of both approaches is to maximize the SV of the ranking values differences, with the weight vector $W$ being optimized for this purpose.	
        \begin{align}\footnotesize	
                 \hspace{-2mm} SV_{TOPSIS} = \sum\nolimits_{i=1}^{N} { \sum\nolimits_{j=i+1}^{N} 
                 {    \lvert  {  \frac{S^-_i}{S^-_i + S^+_i }   -   \frac{S^-_j}{S^-_j + S^+_j }      } \rvert  }    },
                \label{C2}
        \end{align}
        \begin{align}\footnotesize	
                \hspace{-2mm}SV_{SAW} = \sum\nolimits_{i=1}^{N} { \sum\nolimits_{j=i+1}^{N} 
                 {    \lvert  {  \sum\nolimits_{k=1}^{N} { r_{ik}W_k}   -   \sum\nolimits_{k=1}^{N} { r_{jk}W_k}      } \rvert  }    }.
                \label{C3}
        \end{align}

As defined in~\cite{mefgouda2023new}, $(S^+_i)$ represents the euclidean distances between each alternative and the ideal and negative solutions, $(S^-_i)$ refers to the euclidean distance between candidate $i$ and the non-ideal solution. However, $W$ and $r$ refers to the vector of weights and the normalized matrix, respectively.

 \paragraph{Comprehensive Weights} After calculating  $W^S_j$ and $W^O_j$, we set $\alpha=0.2$ and $\beta=0.8$, instead of $\beta=\alpha=1/2$, because the objective weights are more realistic and they reflect the importance of the decision criteria better.
 

\subsubsection{Ranking} According to the suitable decision criteria weights obtained from the precedent step, a MADM approach is used to Rank the available RATs and select the best one.

\section{Experimental Results}\label{Simulation_and_Results}
\subsection{Simulation Settings}


We evaluate the performance of the proposed technique in reducing RRP through MATLAB by creating digital models to simulate HWNs. We then assess its ability to meet service requirements. The simulated HWN environment comprises four types of RANs—WiFi, WiMAX, LTE, and 5G characterized by six attributes: CB, DR, S, D, J, and PLR, as defined in Table~\ref{Attributes_values_for_the_candidate_networks}.

\begin{table}[htp!]
\renewcommand{\arraystretch}{1.3}
\caption{Attributes values for the candidate networks.}
\label{Attributes_values_for_the_candidate_networks}
\centering
\begin{tabular}{p{0.7cm}p{0.9cm}lp{1cm}llp{0.7cm}}
\hline
\bfseries Tech. & \bfseries CB (byte) &  \bfseries S (\%)  & \bfseries DR (mbps) &  \bfseries D (ms) &   \bfseries J (ms) & \bfseries PLR (\%)  \\
\hline
\bfseries WIFI &  	[5-10]&	50&	1-11&	100-150&	10-20&	20-80\\ \hline
\bfseries WiMax	&  [40-50]&	60&	1–60&	60–100&	3–10&	20–80\\ \hline
\bfseries LTE &	[40-50]	& 60	& 2-100	& 50–300	&3–12	&20-80 \\   \hline
\bfseries 5G &	90	& 70	& 400-$10^3$	& 1–10	&1–3	&5-20 \\ 
\hline
\end{tabular}
\end{table}
\begin{figure}[hbt!]
  \centering
  \begin{subfigure}{1\linewidth}
\centering
    \includegraphics[width = 1\linewidth]{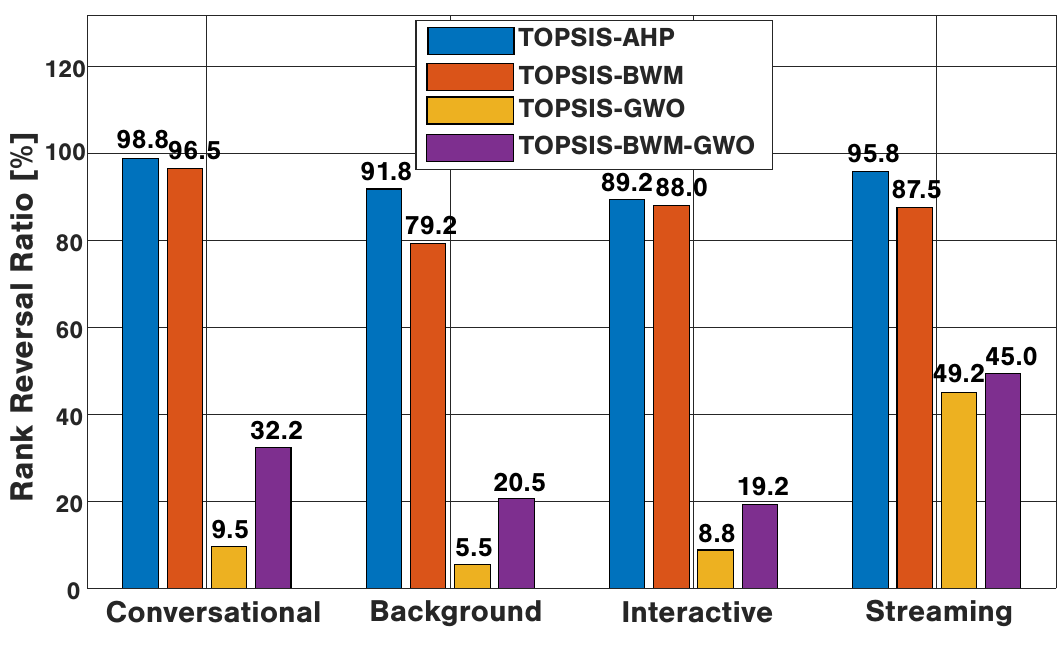}
    \caption{TOPSIS}
  \end{subfigure}%
  \\
  \begin{subfigure}{1\linewidth}
    \centering
    \includegraphics[width = 1\linewidth]{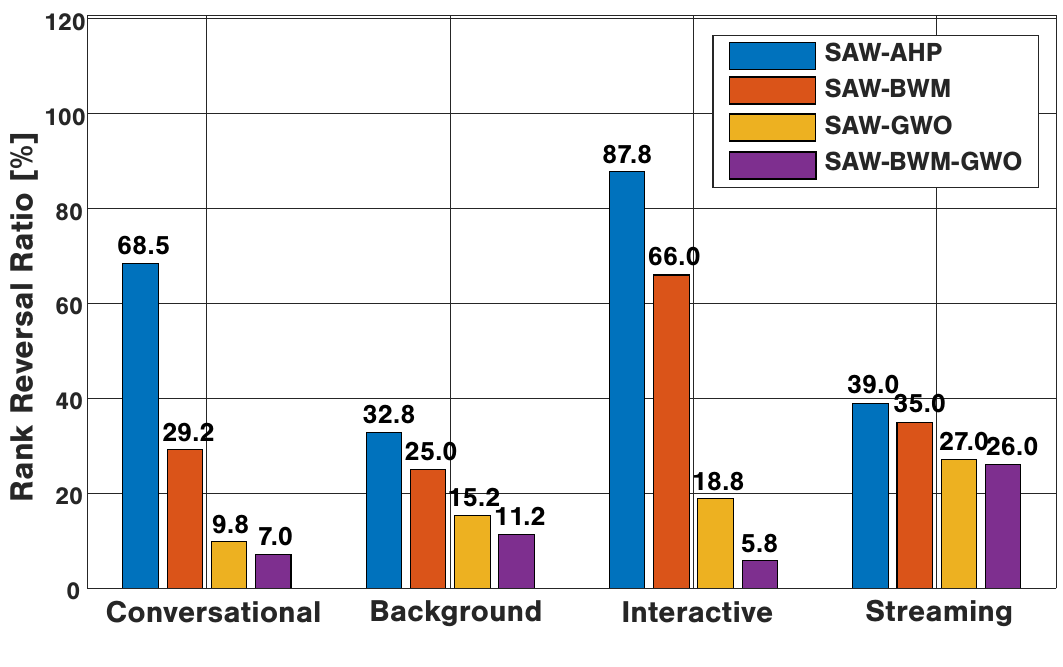}

    \caption{SAW}
  \end{subfigure}%
  \caption{Average rank reversal ratio.}
  \label{fig_RRP}    
\end{figure}


We conducted simulations over \(10^4\) iterations, evaluating and ranking networks. The lowest-performing network was removed at each step, and the algorithm was re-executed. To enhance rank reversal frequency, eight networks were generated per iteration. The proposed approach was compared to traditional AHP with TOPSIS and SAW, assessing the efficiency of TOPSIS-BMW-GWO and SAW-BMW-GWO against TOPSIS-AHP, SAW-AHP, TOPSIS-BWM, and SAW-BWM across four traffic classes.
The resulting AHP weights, used in MADM to assess the relative importance of decision criteria, are presented in Table \ref{AHPTable}~\cite{almutairi2018genetic,mefgouda2023new}.




\begin{table}[hbt!]
\renewcommand{\arraystretch}{1.2}
\caption{Decision criteria weights.}
\label{AHPTable}
\centering
\begin{tabular}{p{2.3cm}p{0.6cm}p{0.6cm}p{0.6cm}p{0.6cm}p{0.6cm}p{0.6cm}}
\hline
\bfseries  & \bfseries CB &  \bfseries S & \bfseries DR &  \bfseries D &   \bfseries J& \bfseries PLR  \\
\hline
\bfseries $W_{conversational}$ &  0.036 &	0.124  & 0.104 & 0.325 &	0.307 &	0.102 \\ \hline
\bfseries $W_{Background}$	&  0.085 &	0.155 &	0.441 &	0.051 &	0.079 &	0.186\\ \hline
\bfseries $W_{Interactive}$& 0.078	& 0.174	&0.092 & 0.309	& 0.050	& 0.294 \\  \hline
\bfseries $W_{Streaming}$ &	0.101	& 0.195	&0.297	& 0.092	&0.119	&0.192 \\  
\hline
\end{tabular}
\end{table}
\vspace{-2mm}
\subsection{Rank Reversal Problem}
The simulation results of the RRP ratio for the TOPSIS-NS and SAW-NS across four traffic classes are shown in Fig.~\ref{fig_RRP}. The comparison of the proposed BWM-GWO method with the AHP method is also shown. The figures illustrate the results for scenarios where the best and worst networks are removed.
The results demonstrate that, although the BWM is a subjective technique, it significantly reduces the RRP when combined with the SAW method compared to the TOPSIS method. The GWO technique enhances the optimization of RRP, while the AHP method is more prone to RRP. Furthermore, the proposed MADM-BWM-GWO outperforms the original AHP weighting method across all traffic classes for both TOPSIS and SAW. Specifically, when applying MADM-BWM-GWO with TOPSIS, the RRP is reduced from 98\% to 32\%, 91.8\% to 20.5\%, 89.2\% to 19.2\%, and 95.8\% to 49\% in conversational, background, interactive, and streaming scenarios, respectively. Similarly, using MADM-BWM-GWO with the SAW approach reduces RRP from 68.5\% to 7\%, 23.8\% to 11.2\%, 87.8\% to 5.8\%, and 39\% to 26\% for these scenarios. {\color{black}These improvements optimize network resource allocation and reduce packet loss, enhancing reliability for latency-sensitive applications.}



\subsection{Service Requiremetns}
\begin{figure}[hbt!]
  \begin{subfigure}{1\linewidth}
\centering 
    \includegraphics[width = 1\linewidth]{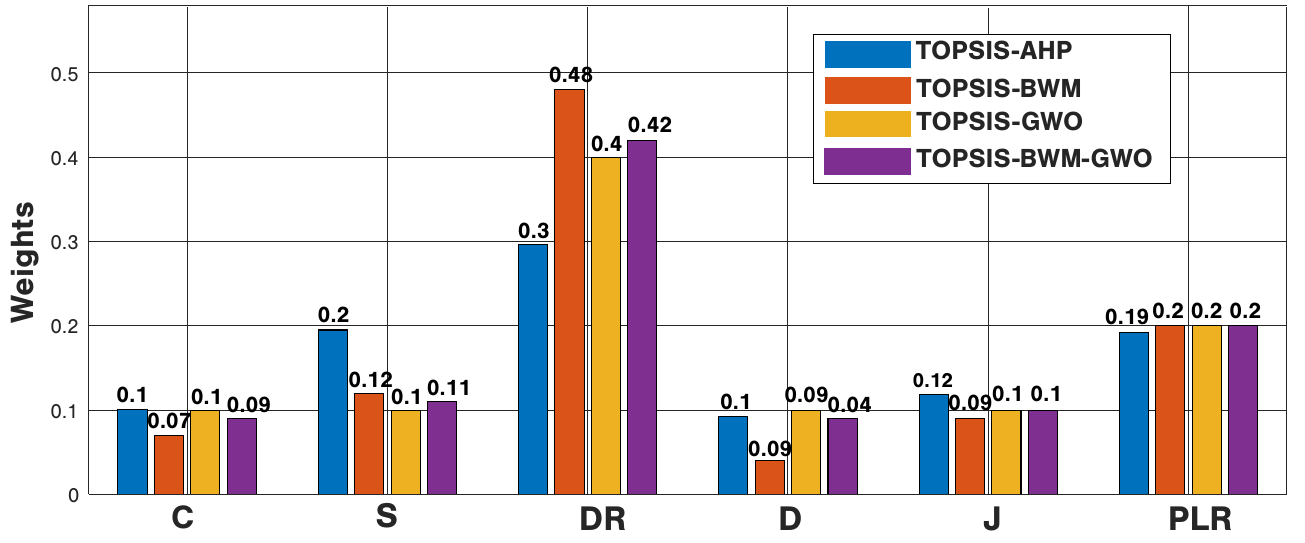}
    \caption{TOPSIS}
  \end{subfigure}%
  \\
  \begin{subfigure}{1\linewidth}

  \centering 

    \includegraphics[width =1\linewidth]{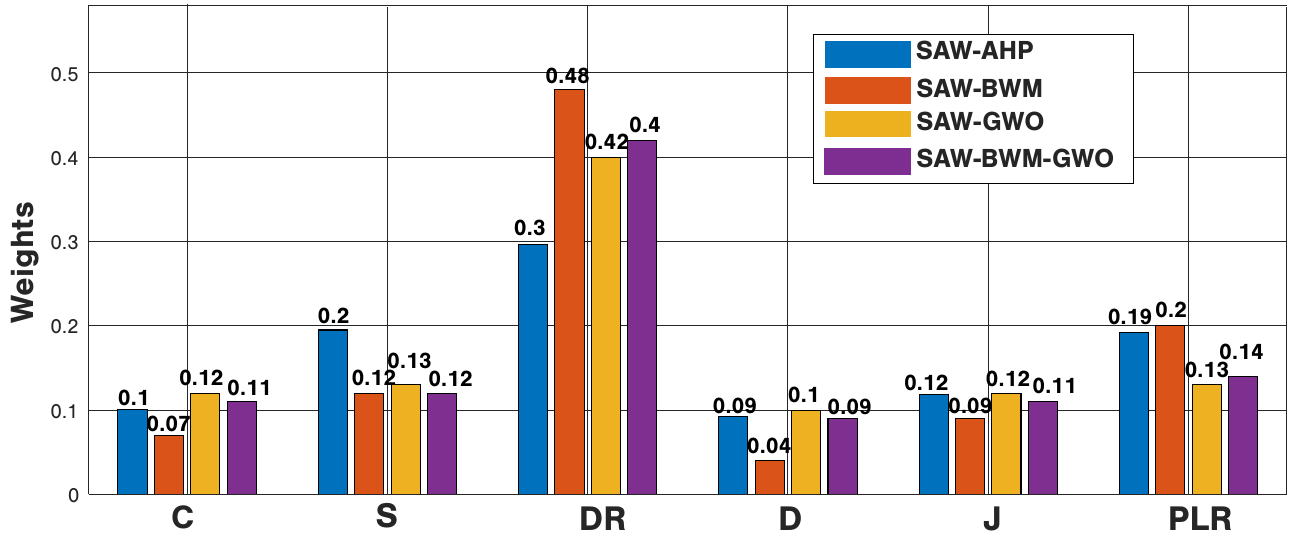}
    \caption{SAW}
  \end{subfigure}%
  \caption{Optimized weight calculation for RAT attributes.}
  \label{fig_service_requir}    
\end{figure}

Fig.~\ref{fig_service_requir} shows the decision criteria weights of the subjective weight, objective weight, and comprehensive weight of attribute indicators under the streaming class and compare them to the AHP technique. 
We observe that the proposed BWM-GWO technique assigns a superior weight value to the DR compared to the AHP technique in both TOPSIS and SAW. This alligns with the service requirements for streaming applications where high DRs and low PLRs are required.
As a result, our proposed MADM-BWM-GWO approach optimizes both the DR and PLR better than the AHP, fulfilling the service requirements. 

\section{Conclusion}
\label{Conclusion}

In this paper we introduced a novel weight assignment method that combines the BWM and GWO to overcome the limitations of MADM-NS. The proposed solution can be integrated with any MADM approach to compute decision criteria weights, reduce the RRP, and meet user/service requirements. Simulation results showed that our approach outperforms the classical AHP technique. The rank reversal problem was reduced for all traffic classes using the TOPSIS and SAW approaches, and all traffic class requirements were satisfied. {\color{black}  For future work, we aim to transform the digital model into a DT for HWNs, enabling dynamic monitoring, predictive analysis, and AI-driven optimization of resources and QoS.}


\end{document}